\documentclass{llncs}

\usepackage{array}
\usepackage{cite} 
\usepackage{url}  
\usepackage{ifthen}  
\usepackage{makeidx}  
\usepackage{multicol}   
\usepackage[utf8]{inputenc} 
\usepackage{amsmath,amssymb}
\usepackage[english]{babel}
\usepackage{graphicx}
\usepackage{hyperref}
\usepackage{verbatim}
\usepackage{color} 
\usepackage{placeins}

\newenvironment{gml}{}{}

\newcommand{\GGLrelease}[0]{GGL v4.0}

\def\startGML{\begin{center}\begin{minipage}{0.9\textwidth}\footnotesize}
\def\endGML{\end{minipage}\end{center}}

\newcommand{\etal}{\textit{et al.}}

\newcommand{\YD}[0]{$\text{Y}$-$\Delta$}
\newcommand{\AGG}[0]{\textsc{AGG}}
\newcommand{\GrGen}[0]{\textsc{GrGen.NET}}
\newcommand{\OpenBabel}[0]{\textsc{OpenBabel}}
\newcommand{\toyChem}[0]{\texttt{toyChem}}
\newcommand{\chemruleSvg}[0]{\texttt{chemrule2svg.pl}}
\newcommand{\printReactionNetwork}[0]{\texttt{printReactionNetwork.pl}}

\newcommand{\NodeSet}{V}
\newcommand{\Node}{v}
\newcommand{\EdgeSet}{E}
\newcommand{\Label}[1]{l_{#1}}
\newcommand{\NodeLabel}{\Label{\NodeSet}}
\newcommand{\EdgeLabel}{\Label{\EdgeSet}}

\newcommand{\RuleLeft}{L}
\newcommand{\RuleRight}{R}

\begin{document}

\title{The Graph Grammar Library - a generic
framework for chemical graph rewrite systems}

\author{Martin Mann\inst{1} \and Heinz Ekker\inst{2} \and Christoph
Flamm\inst{2}}

\institute{%
Bioinformatics, Institut for Computer Science, University of
Freiburg, 79106 Freiburg, Germany
\email{mmann@informatik.uni-freiburg.de} 
\and Institute for Theoretical Chemistry,
University of Vienna,
W{\"a}hringerstrasse 17,
1090 Vienna,
Austria
\email{xtof@tbi.univie.ac.at} 
}

\maketitle

\begin{abstract}
Graph rewrite systems are powerful tools to model and
study complex problems in various fields of research. Their successful
application to chemical reaction modelling on a molecular level was shown
but no appropriate and simple system is available at the moment.

The presented Graph Grammar Library (GGL) implements a generic Double Push Out
approach for general graph rewrite systems. The framework focuses on a high
level of modularity as well as high performance, using state-of-the-art
algorithms and data structures, and comes with extensive documentation.
The large GGL chemistry module enables extensive and detailed studies of
chemical systems. It well meets the requirements and abilities envisioned by
Yadav et al. (2004) for such chemical rewrite systems. Here, molecules are
represented as undirected labeled graphs while chemical reactions are described
by according graph grammar rules. Beside the graph transformation, the GGL
offers advanced cheminformatics algorithms for instance to estimate energies
ofmolecules or aromaticity perception. These features are illustrated
using a set of reactions from polyketide chemistry a huge class of natural
compounds of medical relevance.

The graph grammar based simulation of chemical
reactions offered by the GGL is a powerful tool for extensive cheminformatics
studies on a molecular level. The GGL already provides rewrite rules for all
enzymes listed in the KEGG LIGAND database is freely available at \\
\url{http://www.tbi.univie.ac.at/software/GGL/}.
\end{abstract}


\section{Background}

Graphs are powerful tools to represent and study all kind of data in any field
of research. In order to generate graph structures of interest or to alter them
according to some directive, graph transformations can be applied. A common
approach is to formulate such transformations in terms of graph grammars or graph
rewrite systems \cite{Beck:04,Blinov:06,Rudolf:00}. This enables a compact but
very expressive representation of allowed alterations and allows for sound
mathematical analyses of the problems \cite{Andersen:11,Flamm:10}.

Here, we present our Graph Grammar Library (GGL), a fast and generic C++
framework to formulate and apply graph grammars. Beside the general graph
rewrite system, we provide a specialized module to enable efficient graph
grammar applications in chemical context. The power of such systems for
chemical studies was highlighted by Yadav \etal{}~\cite{Yadav:04} who emphasized
the lack and need for an efficient implementation. The requirements and
abilities sketched by Yadav \etal{} are well met by the capabilites of
the GGL framework as presented within this manuscript.

\begin{figure}[tb]
    \begin{center}
    	\includegraphics[width=\textwidth]{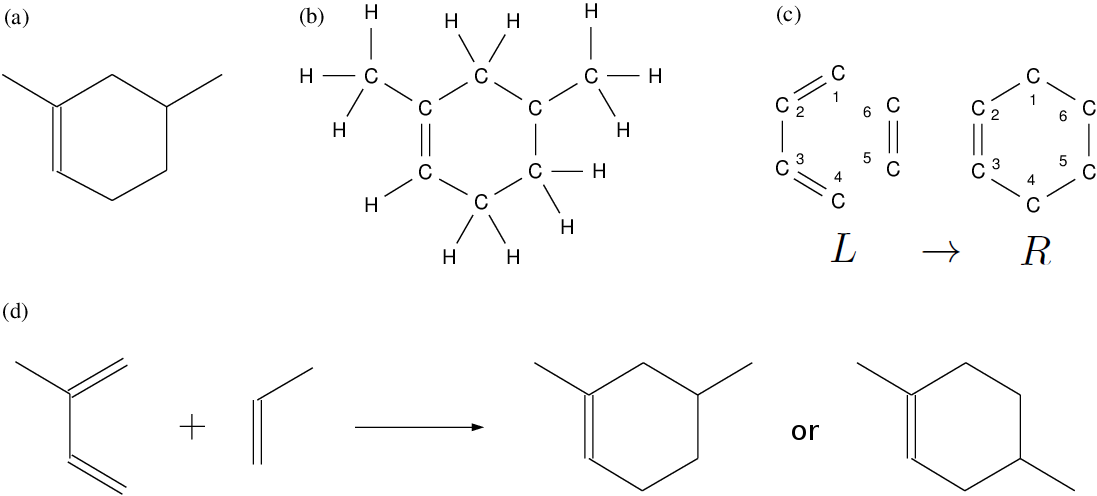}
    \end{center}
	\caption{Two graph representations of the same molecule at different
levels of detail (a, b). Graphical depiction of the Diels-Alder
reaction (d) for two molecules that enables two different result
molecules. (c) The corresponding graph grammar rule
$\RuleLeft\rightarrow\RuleRight$.} 
	\label{fig:reaction-example}
\end{figure}

To represent molecules in graph notation is well known in chemistry and found in
any text book. Therein, atoms are represented by labeled nodes in the graph and
the connecting chemical bonds are depicted by according edges (see
Fig.~\ref{fig:reaction-example}~a) and~b)). Such molecule graphs can be used to
model chemical reactions by the application of graph grammar rules as shown in
Fig.~\ref{fig:reaction-example}~c) and~d). This small example shows already the
power of graph rewrite systems, since a single graph grammar rule encodes all
possible interaction combinations of the chemical reaction encoded (within the
example two possible product molecules are possible, see
Fig.~\ref{fig:reaction-example}~d)).

In the following, we will introduce the concept of graph rewrite systems in
general and how it is implemented in detail within the GGL framework. We give
insight into the techniques applied and the efforts done in order to gain an
efficient and flexible framework for graph grammar applications. A major
contribution of the GGL is its chemistry module tailored to fulfill the needs
when representing chemical reactions via graph transformations as requested by
Yadav \etal{}~\cite{Yadav:04}. We show the generality of the GGL via
several applications and elaborate its use for chemical studies. The latter
is extended by providing a full set of predefined graph grammar rules for all
enzymes listed in the KEGG LIGAND database \cite{Kanehisa:11} (Release 58.1 June
2011).

\subsection*{Graph Grammars and their Applications in Chemistry}

For simplicity, we will focus on labeled, undirected graphs in the following.
Such a graph is given by a tuple $(\NodeSet,\EdgeSet,\NodeLabel,\EdgeLabel)$
that is defined by the set of $n$~nodes $\NodeSet=\{\Node_1,\ldots,\Node_n\}$,
the set of edges $\EdgeSet=\{\{\Node_i,\Node_j\}\;|\;
\Node_i,\Node_j\in\NodeSet\}$, and the label functions $\NodeLabel : \NodeSet
\rightarrow \Sigma^\ast$ and $\EdgeLabel : \EdgeSet \rightarrow \Sigma^\ast$
that assign a label based on some alphabet~$\Sigma$ to each node and edge,
respectively.

Graph rewrite systems are an algebraic approach to apply graph transformations
\cite{Rozenberg:97}. It is defined by a set of rewriting rules of the form
$\RuleLeft\rightarrow\RuleRight$, with $\RuleLeft$ defining the subgraph to be
replaced during the transformation, the \emph{pattern} or \emph{left side} of
the rule, and $\RuleRight$ stating the transformation result, i.e. the
\emph{replacement} or \emph{right side} of the rule. Thus, graph rewriting
requires the location of the pattern graph within the graph to transform. This
\emph{subgraph monomorphism problem} is a hard computational problem depending
on the graph type \cite{Eppstein:95}. Formally, given two graphs
$G=(\NodeSet^G,\EdgeSet^G,\NodeLabel^G,\EdgeLabel^G)$ and
$\RuleLeft=(\NodeSet^\RuleLeft,\EdgeSet^\RuleLeft,\NodeLabel^\RuleLeft,\EdgeLabel^\RuleLeft)$,
one has to find an injective mapping $m:\NodeSet^\RuleLeft
\rightarrow\NodeSet^G$ such that it holds
\begin{eqnarray*}
\forall_{\Node_i,\Node_j\in\NodeSet^\RuleLeft} &:& m(\Node_i) \neq m(\Node_j)
\quad\wedge\quad \NodeLabel^\RuleLeft(\Node_i) = \NodeLabel^G(m(\Node_i))
	\\
\forall_{\{\Node_i,\Node_j\}\in\EdgeSet^\RuleLeft} &:& \{m(\Node_i),m(\Node_j)\}
\in \EdgeSet^G  \quad\wedge\quad \EdgeLabel^\RuleLeft(\{\Node_i,\Node_j\}) =
\EdgeLabel^G(\{m(\Node_i),m(\Node_j)\})
\end{eqnarray*}
In the following, such a mapping~$m$ is called a \emph{match} of the
pattern~$\RuleLeft$ within target graph~$G$ and defines the subgraph of~$G$
isomorph to~$\RuleLeft$.

When applying a graph grammar rule $\RuleLeft\rightarrow\RuleRight$, one can
follow the Double Push Out (DPO) approach \cite{Rozenberg:97,Corradini:97}.
Therein, given a match~$m$ of~$\RuleLeft$ in some target graph~$G$, the result
graph~$G'$ is derived from~$G$ by (I) a relabeling of all matched nodes present
in both $\NodeSet^\RuleLeft$ and $\NodeSet^\RuleRight$ but showing different
labels according to $\NodeLabel^\RuleLeft,\NodeLabel^\RuleRight$, (II) the
deletion of all nodes (and adjacent edges) present only in $\NodeSet^\RuleLeft$
but absent in $\NodeSet^\RuleRight$, and (III) the adding of all nodes
exclusively present in $\NodeSet^\RuleRight$. Edges are handled accordingly
based on $\EdgeSet^\RuleLeft$ and $\EdgeSet^\RuleRight$. For further details
please refer to the standard literature, e.g. \cite{Rozenberg:97,Corradini:97}.

Such graph grammar rules have been successfully applied to model chemical
reactions with molecular detail \cite{Benkoe:03,Flamm:10,Rossello:05}. Therein,
a graph grammar rule encodes the molecule graph transformations resulting from a
chemical reaction as exemplified in Fig.~\ref{fig:reaction-example}. The figure
shows the power of such an encoding: the reaction is as much abstracted from the
specific molecules as possible and implicitely encodes all interaction
configurations.

At a higher abstraction level, frameworks have been introduced to encode and
alter metabolic and signal-transduction networks based on graph rewrite systems
\cite{Blinov:06,Colvin:10,Danoes:07,Danos:08,Hlavacek:06,Xu:11}. The focus here
is to encode the kinetics and interactions of chemical reactions and not the mechanisms
underlying them.

All the above mentioned frameworks are either prototypical implementations not
tailored for a extensive application or very specialized systems designed for a
specific type of experiment/problem. Within the context of general graph rewrite
systems, other implementations have been introduced. Among them are the \AGG{}
\cite{Taentzer:00,Taentzer:04} and the \GrGen{} \cite{Geiss:06,Jakumeit:10}
frameworks. While the \AGG{} implementation was shown to be not well performing
for larger data sets \cite{Geiss:06}, we found the very general \GrGen{} system
not well placed to meet the requirements formulated by Yadav
\etal{}~\cite{Yadav:04}. The compilation of the graph grammar rules into
executable code including the planned search strategy used to locate the
patterns and the very expressive but complicated rule encoding makes the package
powerful but not well placed for its integration into tools for chemical
modelling to be used by chemists.

Here, we introduce the Graph Grammar Library (GGL) as a generic graph grammar
framework with a strong focus on chemical applications as sketched by Yadav
\etal{}. The GGL is not as general as e.g. the \GrGen{} system, since it
currently allows only for a single label attribute associated to each node and
edge, but it features an easy rule encoding in combination with a flexible and
efficient rule application framework. In the following, we will introduce the
GGL and the applied methods in detail.

\section{The Graph Grammar Library}

The Graph Grammar Library (GGL) is a generic C++ programming library that
enables an easy setup of graph rewrite systems for labeled graphs. During the
design of the library, we have focused on the following features:

\begin{description}
  \item[Generality] The GGL is built as a generic framework to be used for
  graph grammars on labeled graphs. The implementation uses the generic Boost Graph Library
  (BGL) \cite{Siek:01} for its core graph representations. This enables the
  embedding of the graph rewrite functionalities into existing BGL-based
  projects. Our chemistry module supports the full atom range as well as
  standard chemoinformatics formats and libraries \cite{OBoyle:11,Weininger:88}
  \item[Modularity] The object-oriented modular design of the library enables a
  clear separation of functional units and the straight-forward implementation
  and use of specific functionalities. The interconnection of the modules is
  defined by clear and slim interfaces to enable a high level of transparency.
  The separation into context-specific sublibraries enables a selective use of
  the operations needed for a specific task.
  \item[Performance] The main goal of the library is to provide a generic
  framework for computationally extensive applications. Thus, the implementation
  is tuned to be as fast as possible while maintaining a high level of
  generality and modularity. We apply efficient state-of-the-art algorithms
  \cite{Cordella:04,Guzman:03,Hanser:96} and data structures
  \cite{Costa:10,Siek:01}, as detailed later, and have profiled and improved the
  code for a maximal performance.
  \item[Documentation] In order to make the GGL easily applicable, we provide
  a well documented Application Programming Interface (API) for programmers as
  well as tutorials for end users focusing e.g. on the graph grammar rule
  encoding.
\end{description}

\noindent In the following, we will present the core functionalities of the
library and how graph grammar rules are formulated and applied. Afterwards, we
focus on the chemistry module of the GGL and the provided features and
functionalities for chemoinformatical applications.

\subsection{Subgraph Matching}

As introduced above, the identification of matches~$m$ of a given left side
pattern~$\RuleLeft$ of a graph grammar rule $\RuleLeft\rightarrow\RuleRight$ is
the central task of rule applications. The applied algorithm to solve the
subgraph isomorphism problem is thus defining the performance of the whole
rewrite engine~\cite{Geiss:06}. Within the GGL, we apply the efficient
\texttt{VF2}-algorithm and implementation introduced by Cordella
\etal{}~\cite{Cordella:04}, which is among the fastest
available~\cite{Cordella:99}. We slightly adapted and extended the fast
C-implementation and provide it within the \GGLrelease{} package.

We have extended the implementation in two directions. First, we have introduced
the handling of wildcard labels. This is needed to specify the existence of a
given node without specifying the concrete label, e.g. to define an adjacent
residual group of a molecule without any details. Furthermore, we have added an
advanced constraint handling that can be used to enforce additional matching
constraints. Among them are degree and adjacency constraints, negative
application conditions (NAC), like the non-existence of an edge, as well as
advanced chemistry related confinements.

Since we focused on a high level of modularity, we use a clear interface to
provide the subgraph matching functionalities within the GGL. The
\texttt{VF2}-algorithm is ported via this interface for its application. This
enables the use of the available efficient \texttt{VF2}-implementation as well
as the replacement of the \texttt{VF2}-algorithm with other subgraph matching
approaches if needed. Since it was shown that the isomorphism problem can be
solved efficiently for some types of graphs \cite{Eppstein:95}, it might be
useful to apply a dedicated matching algorithm depending on the problem at hand.
Furthermore, other matching approaches can be easily integrated
\cite{Doerr:95,Rudolf:00,Varro:06}. The whole subgraph matching module is
encapsulated into an independent library module, the SubGraph Matching (SGM)
library, which is part of the GGL distribution.

\subsection{Rule Encoding}

Before we give details on the graph grammar rule applications we first introduce
how rules are represented and to be specified within the GGL framework. Within
the library, a graph grammar rule is represented by a specific graph object that
encodes both, the left ($\RuleLeft$) and right side part ($\RuleRight$) of a
rule $\RuleLeft\rightarrow\RuleRight$. Thus, it holds for each node and edge if
it is present in $\RuleLeft$ or $\RuleRight$ together with the according label.
Formally, it is defined by the extended graph tuple $(\NodeSet,\EdgeSet,
\NodeLabel^{\RuleLeft},\NodeLabel^{\RuleRight},
\EdgeLabel^{\RuleLeft},\EdgeLabel^{\RuleRight})$, which encodes the mapping of
left and right side graphs. A node $\Node\in\NodeSet$ \emph{not present} within
$\RuleLeft$ will have an empty left side node
label~$\NodeLabel^{\RuleLeft}(\Node)=\lambda$. The same holds for edges. Thus we
can derive the left side pattern $\RuleLeft =
(\NodeSet^\RuleLeft,\EdgeSet^\RuleLeft,
\NodeLabel^\RuleLeft,\EdgeLabel^\RuleLeft)$ by
\begin{eqnarray*}
	\NodeSet^\RuleLeft = \{\Node\in\NodeSet \;|\;
	\NodeLabel^{\RuleLeft}(\Node)\neq\lambda \} 
	&\text{ and }&
	\EdgeSet^\RuleLeft = \{\;\{\Node_i,\Node_j\}\in\EdgeSet \;|\;
	\EdgeLabel^{\RuleLeft}(\{\Node_i,\Node_j\})\neq\lambda \}
	.
\end{eqnarray*}
\noindent The right side graph~$\RuleRight$ is derived accordingly. Based on
this representation, we have all the information at hand to apply the rule for each
match~$m$ of~$\RuleLeft$ within some target graph~$G$, since we know exactly the
corresponding nodes/edges of~$\RuleLeft$ and~$\RuleRight$.

For an easy and readable specification of graph grammar rules, we use an
encoding in the Graph Modelling Language (GML) format \cite{Himsolt:99}. This
key-value pair structured format enables a compact and human-readable encoding
while it is still machine-parsable due to its simple grammar.
Figure~\ref{gml:Diels-Alder} shows the GML encoding of the rule presented in
Fig.~\ref{fig:reaction-example}. Note, all unchanging nodes and edges part of
$\RuleLeft$ \emph{and} $\RuleRight$ are defined within the \texttt{context}
section (contributing to both $\NodeLabel^{\RuleLeft}$ and
$\NodeLabel^{\RuleRight}$), while left/right side graph specific nodes/edges and
their labels are given in the according \texttt{left}/\texttt{right} sections
and will only define the according $\NodeLabel^{\RuleLeft}$/
$\NodeLabel^{\RuleRight}$ data, respectively.

\begin{figure}[tb] 
 
\begin{center}
\begin{gml}
\begin{minipage}{0.7\textwidth}
\tiny
\makebox[\textwidth]{\hrulefill}
\begin{verbatim}
rule [
  ruleID "Diels-Alder reaction"
  context [
    node [ id 1 label "C" ]
    node [ id 2 label "C" ]
    node [ id 3 label "C" ]
    node [ id 4 label "C" ]
    node [ id 5 label "C" ]
    node [ id 6 label "C" ]
  ]
  left [
    edge [ source 1 target 2 label "=" ]
    edge [ source 2 target 3 label "-" ]
    edge [ source 3 target 4 label "=" ]
    edge [ source 5 target 6 label "=" ]
    constrainNoEdge [ source 1 target 5 ]
    constrainNoEdge [ source 4 target 6 ]
  ]
  right [
    edge [ source 1 target 2 label "-" ]
    edge [ source 2 target 3 label "=" ]
    edge [ source 3 target 4 label "-" ]
    edge [ source 4 target 5 label "-" ]
    edge [ source 5 target 6 label "-" ]
    edge [ source 6 target 1 label "-" ]
  ]
]
\end{verbatim} 
\makebox[\textwidth]{\hrulefill}
\end{minipage}
\end{gml}
\end{center}
	\caption{The GML rule encoding of the graph grammar rule presented in
Fig.~\ref{fig:reaction-example}.}
	\label{gml:Diels-Alder}
\end{figure}

If the label of some nodes/edges of the pattern is of no interest for the
matching, one can specify a wildcard label (e.g. *) by adding the according
key-value entry \fbox{\texttt{wildcard "*"}}. The subgraph matching engine
will now match all nodes/edges from the left side pattern~$\RuleLeft$ that show
the wildcard label, e.g.
\fbox{\texttt{node $[$ id 1 label "*" $]$}},
to any other nodes/edges in the target graph without further label
comparisons.

Often, the specification of wildcard labels makes the rule too general and might
enable unintended rule applications. To tackle this problem, we support the
specification of additional matching constraints that have to be ensured by the
matching procedure. The simplest example is to restrict the allowed labels for a
node defined with wildcard label to a given set of labels:
\begin{center}
\fbox{\texttt{constrainNode [ id 1 op = nodeLabels [ label "C" label "N" ] ]}}
\end{center}

\noindent The reverse constraint can be encoded by changing the operator value
from ``\texttt{=}'' to ``\texttt{!}'' to disallow the given labels. Other
constraints supported for general graph grammars are the restriction of edge
labels or node degree, the specification of required/forbidden adjacent nodes or
edges, or the explicit prohibition of the existance of an edge between two
nodes. For further details on the constraints supported and their encoding we refer to
the according tutorial part of the \GGLrelease{} distribution.

\subsection{Rule Application}

The rule application follows the Double PushOut (DPO) approach
\cite{Rozenberg:97,Corradini:97}. Given a graph grammar rule
$\RuleLeft\rightarrow\RuleRight$ encoded by a rule graph $(\NodeSet,\EdgeSet,
\NodeLabel^{\RuleLeft},\NodeLabel^{\RuleRight},
\EdgeLabel^{\RuleLeft},\EdgeLabel^{\RuleRight})$ as defined above and a target
graph~$G$ onto which the rule is to be applied. First, all matches~$m$ of the
rule's left side pattern~$\RuleLeft$ within~$G$ have to be indentified. The GGL
uses to this end its subgraph matching interface and (as default) the
\texttt{VF2}-algorithm~\cite{Cordella:04} as already introduced. For each
match~$m$, the following procedure is applied to generate the result graph~$G'$
for the current match and rule:

\begin{enumerate}
  \item Remove from~$G$ all nodes exclusive within~$\RuleLeft$
  \begin{eqnarray*}
  \NodeSet^{G'} &=& \NodeSet^G \setminus \{ m(\Node) \;|\; \Node\in\NodeSet
  \wedge \NodeLabel^\RuleRight(\Node) = \lambda\} 
  \end{eqnarray*}
  and all edges exclusive within~$\RuleLeft$ or adjacent to removed nodes.
  \begin{eqnarray*}
  \EdgeSet^{G'} = \EdgeSet^G \setminus &\{& \{m(\Node_i),m(\Node_j)\}
  \;| \;\;\{m(\Node_i),m(\Node_j)\}\in\EdgeSet\\
   &&\; \wedge\;
  (\EdgeLabel^\RuleRight( \{m(\Node_i),m(\Node_j)\} ) = \lambda
  \; \vee \; \NodeLabel^\RuleRight(\Node_i) = \lambda
  \vee \NodeLabel^\RuleRight(\Node_j) = \lambda) \;\}
  \end{eqnarray*}
  \item Relabel within~$G'$ all nodes/edges showing different (non-empty) labels
  within $\RuleLeft$ and $\RuleRight$.
  \item Add to~$G'$ all nodes/edges exclusive within~$\RuleRight$ and label
  accordingly.
\end{enumerate}

All resulting graphs~$G'$ are then forwarded to a provided instance of a well
defined graph reporter interface. This enables application specific
post-processing and storing of the graphs resulting from the graph rewrite.
Within the chemistry module, described next, this is for instance used to apply
sanity checks on the produced molecule graphs, to convert and store molecule
graphs into canonical string formats, to gather reaction product information, or
to compute the reaction rate of the specific reaction defined by the rule
application. In a more algorithmic context, such a graph reporter can also
trigger another recursive iteration of rule applications resulting in a
depth-first search/traversal of the graph space encoded by the graph grammar. An
according generic implementation is provided and applied within the GGL
framework.


\subsection{Chemistry Package}

As discussed in the introduction, chemistry and especially chemical
reactions are a well placed target for the application of graph rewrite
systems~\cite{Benkoe:03,Rossello:05}. The GGL provides a specialized
chemistry module tailored for such needs. It enables a flexible but fast
and efficient implementation of graph grammar based algorithms to solve
chemical reaction related problems. In the following, we will present the
features provided by the module and the available data structures and
algorithms.

\subsubsection*{Molecules as Graphs}

Figure~\ref{fig:reaction-example} illustrate the
definition of molecules in terms of graphs and the relation of chemical
reactions and graph grammar rules. In detail, a molecule is represented as a
graph~$M$ where the node ($\NodeLabel^M$) and edge labels ($\EdgeLabel^M$) are
restricted to the existing atom and bond types, respectively. As suggested by
Yadav \etal~\cite{Yadav:04} and applied in literature
\cite{Benkoe:03,Rossello:05}, we follow the SMILES encoding of atoms and bonds
\cite{Weininger:88}. Therein, bond labels are confined to
$\EdgeLabel^M(..)\in\{\text{\texttt{-,=,\#,:}}\}$, encoding for single, double,
triple, or aromatic bonds respectively. Atom labels $\NodeLabel^M$ are either
one of the labels known from the periodic table of elements, e.g. \texttt{C} or
\texttt{Br}, an organic atom participating in an aromatic ring (encoded by lower
case symbols, e.g. \texttt{c} or \texttt{n}), or a ``complex'' label encoding
additional charge information, e.g. \texttt{O-}.

Within the GGL chemistry module we enforce an explicit encoding of hydrogen
information as atoms (see Fig.~\ref{fig:reaction-example}b)), even if the SMILES
notation allows for their encoding within complex node labels. This was done a)
to ensure explicit and complete chemical rule encodings, b) to enable sanity
checks for molecules and rules, c) to allow for the prediction of aromaticity
information, and d) to ensure an efficient matching of chemical rule patterns
onto molecule graphs, which is needed for their application.

Based on our internal molecule graph representation, the GGL chemistry module
features a couple of functionalities needed within the context of chemical
reaction problems. This includes sanity checks for molecules (e.g. correct label
usage, valence constraints, etc.), the SMILES string encoding (discussed within
next section), and automatic hydrogen prediction. Furthermore, we have
reimplemented the efficient free energy prediction algorithm introduced
by Jankowski \etal{}~\cite{Jankowski:08}. It enables a fast energy approximation
of a given molecule graph based on a decomposition into defined atomic groups
and their energy contributions. Again, we take advantage of the fast subgraph
matching module part of the \GGLrelease{} to allow for a performant
decomposition. The implementation and use of other energy estimation approaches
is easily possible due to the modular design of the library.

The estimation of a molecule's energy usually requires knowledge about (hetero)
aromatic rings within the molecule~\cite{Schleyer:96}. Since the property of
aromaticity might emerge or vanish due to chemical reactions, we provide an
aromaticity prediction framework. It is based on a ring perception using an
extension of the fast algorithm suggested by Hanser \etal{}~\cite{Hanser:96} in
combination with a support vector machine learning approach for the aromaticity
prediction based on the NSPDK graph feature kernel~\cite{Costa:10}.

\subsubsection*{Canonical SMILES}

The SMILES notation was actually introduced to enable a string representation of
chemical molecules, i.e. SMILES strings are a string encoding of molecule
graphs~\cite{Weininger:88}. Furthermore, they allow for a canonical, i.e.
unique, representation of molecules~\cite{Weininger:89}. This is especially
useful when storing molecule information in databases or when molecules are to
be given as program input (see \cite{Yadav:04}). The GGL features a full-fledged
canonical SMILES writer implementation as well as an according SMILES parser to
enable SMILES as the chemical communication language for applications.
Internally, molecules are represented as graphs as given above.

\subsubsection*{Chemical Reactions}

Chemical reactions are special graph grammar rules since they have certain
essential side constraints to be fulfilled. First and most important:
conservation of mass, i.e. no atom can appear or vanish. Thus, a chemical
reaction is -- as a coarse grained sketch -- a ``rewiring'' of bonds within or
between molecule graphs. Technically, we can encode chemical reactions as graph
grammar rules in GML notation as introduced above. Therein, atom and node labels
are restricted to the encoding supported by SMILES (see above), while wildcard
labels are allowed to enable more general encodings. All chemical rules can be
and are checked for their correctness within the GGL chemistry framework. Among
the tests are checks for mass conservation, label use, or reasonable valence
changes. Another important check is to ensure that no bond is formed twice,
which is done by implicitely adding according ``no-edge'' constraints for all
bonds formed within the chemical rule.

When applying chemical rules on molecule graphs we can use the generic graph
grammar rule application framework described above. No adjustments for the rule
applications are needed. Since some chemical rewrites might result in
non-realistic molecules, e.g. due to steric constraints not covered by the
rule~\cite{Yadav:04}, we provide a post-application verification step. Here, the
output molecules are postprocessed, e.g. to correct the molecules aromaticity,
and checked for sanity. A chemical rewrite only results in a chemical reaction,
if all result molecules have been shown to be valid. The canonical SMILES
encoding introduced above is used to derive a compact string representation of
both the resulting molecules as well as the whole reaction.

Chemical reaction networks are often subject of reaction pathway analyses. For
this purpose one has to know or estimate the reaction rates to be associated to
individual chemical reactions e.g. as produced by a graph grammar rule
application. The GGL chemistry framework fully supports such requirements. Based
on the approach by Jankowski \etal{}~\cite{Jankowski:08}, we estimate the energy
difference~$\Delta{E}$ of the input and output molecules of a reaction. This
enables the estimate of the reaction rate using the well known Arrhenius law,
i.e. the reaction rate is approximated by $\exp^{-\Delta{E}/RT}$ for a given
temperature~$T$ using the gas constant~$R$. Other approaches, e.g. the machine
learning approach by Kayala \etal{}~\cite{Kayala:11}, can be easily integrated
and used within the GGL chemistry framework if needed due to its modular
architecture.

\subsubsection*{Molecular Group Specification}

The specification of (bio)chemical reactions often requires the representation
of large (unchanged) parts of molecules in order to make the rule as specific as
the chemical reaction. A classic example is the involvement of helper molecules
like ATP, NADH, etc. that are only slightly changed but have to be
represented completely to avoid the application of the rule using similar molecules.

To this end, the GGL supports the specification of molecular groups as
pseudo-atoms within chemical rule definitions. They allow for a much
easier and compact rule definition and avoid potential typos and mistakes.
Section~B of the suppl. material exemplifies the problem.

\subsubsection*{Visualization}

To enable an easier definition and evaluation of chemical reaction data,
visualization scripts are provided. 2D-layouting of molecule graphs is done via
the \OpenBabel{} framework~\cite{OBoyle:11} and scalable vector graphics in SVG
or PDF format are generated.
For instance, given a valid GML encoding of a chemical rule, the script
\chemruleSvg{} can be used to generate an according depiction.
Figure~\ref{fig:Diels-Alder} (suppl. material) exemplifies the application
for the GML rule encoding given in Fig.~\ref{gml:Diels-Alder}.

\begin{figure}[tb] 
 
    \begin{center}
    	\includegraphics[width=1.0\textwidth]{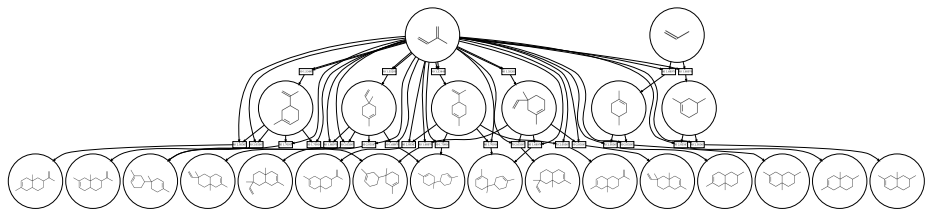}
    \end{center}

	\caption{The depiction of the Diels-Alder reaction network for the two input
molecules depicted in Fig.~\ref{fig:reaction-example}d) (with SMILES
\texttt{C=CC(C)=C} and \texttt{C=CC}) and the rule encoding from
Fig.~\ref{gml:Diels-Alder} using two rule application iterations. 
}
	\label{fig:Diels-Alder:RN}
\end{figure}

The \GGLrelease{} package provides a reimplementation of the reaction network
expansion approach presented in~\cite{Benkoe:03} named \toyChem{}. Given a set
of molecules and chemical reactions, \toyChem{} expands the according reaction
network via an iterative application of the reaction rules. Along the expansion,
it automatically computes according reaction rates and produces a graph encoding
of the reaction network. The script \printReactionNetwork{} visualizes
the network including depictions of the molecules and rate information.
For an example see Fig.~\ref{fig:Diels-Alder:RN}.

\subsubsection*{Set of Enzymatic Reaction Data Provided}

For the application of the GGL to simulations of biochemical reactions within
metabolic networks, 5133 unique GML-encoded graph grammar rules for enzymatic
reactions are included in the \GGLrelease{} distribution. The rules are derived
from the KEGG LIGAND database \cite{Kanehisa:11}, Release 58.1 June 2011. 

For each enzyme listed in the database, a rewrite rule for each reaction
annotated to that enzyme has been created automatically. To this end, an
atom-to-atom correspondence between the substrate and product molecules has been
determined to identify the broken and formed bonds along the given reaction. The
atom mapping was generated using a greedy heuristic losely based on the Cut
Successive Largest algorithm proposed in \cite{Crabtree:09}. The predicted
mappings follow the Principle of Minimal Chemical Distance, which states that
the mechanism involving the least reconfigurations of valence electrons
(i.e.~edge additions/removals) is most likely the true reaction mechanism.
Reactions that are unbalanced or contain compounds with missing or faulty
molecular structure data cannot be mapped and are therefore not included. Since
there is no data on reaction reversibility in the database, only the forward
direction as given in the database is represented.

\subsubsection*{\OpenBabel{} Port}

The GGL chemistry module is of course not the only available cheminformatics
package on the market. It is neither intended nor designed to enable very
sophisticated chemical informations as molecule mass or to provide specific
tools like graphical depiction algorithms already available in other libraries.
A powerful and freely available library is provided by the \OpenBabel{} project
\cite{OBoyle:11} that was initially started to enable an easy conversion between
the various chemical data formats. Nowadays, various tools and solutions are
provided to tackle chemoinformatics problems.

Since we focus on a highly modular design of the GGL, we provide an easy port to
convert our internal molecule graph representation into an \OpenBabel{}~object.
This port can be used to get access to the full set of functionality provided by
the \OpenBabel{} library if needed \cite{OBoyle:11}.

\subsection{Further Library Features}

Beside the features introduced we want to give some further remarks on the GGL
programming library. The object-oriented C++ source code is fully
\texttt{ANSI}-conform and extensively documented. This enables the generation of
the provided Application Programming Interface using the \texttt{doxygen}
system. The dependency checking and compilation process is tailored for
\texttt{Unix}-like systems (including \texttt{Cygwin} for MS Windows) and is
based on the established \texttt{GNU} \texttt{autotools} resulting in an easy
and automatic setup and installation of the libraries and tools. The
\GGLrelease{} comes with an extensive test set framework to ensure the
correctness of the build on the used platform and to ensure the stability and
maintainability of the library. In addition, we provide \texttt{Perl-5} bindings
to enable the application of the GGL functionalities within fast prototypical
\texttt{Perl} developments, e.g. in combination with the \texttt{PerlMol}
package~\cite{TubertBrohman:04}.

Last but not least, the \GGLrelease{} package is freely available at:
\begin{center}
	\url{http://www.tbi.univie.ac.at/software/GGL/}
\end{center}

\section{Application and Examples} \label{sec:application}

The graph grammar library is designed to support a wide range of graph
rewrite systems on any type of graph. For illustration, we have implemented
a few example applications that cover different aspects of graph
rewrite. All are distributed within the \GGLrelease{} package and described
in Sec.~C of the suppl. material. Since the \GGLrelease{} supports an
extensive module for chemical graph rewrite systems, we will focus on
chemical applications in the following.


The graph grammar approach to chemical transformation gives a very compact
description of a whole ``chemical universe'', i.e.\ the language of all
chemical graphs, which are reachable from an initial set of chemical
``starting graphs'' by iterative application of the reaction rules. The
iterative expansion of a particular graph grammar yields a directed,
potentially infinit reaction network, where the chemical graphs are
connected by hyperedges representing the reaction rules. These reaction
networks can be further analysed statistically or with methodologies from
network theory to uncover unexpected relations between network nodes and
their properties \cite{Grzybowski:2009,Soh:2012}. The \toyChem{}
utility distributed with the GGL implements only simple exhaustive
iterative expansion of a chemical universe. For the efficient exploration
of a chemical universe a sofisticated strategy framework is required (see
\cite{Andersen:2013}) to avoid combinatorial explosion. The GGL is a major
improvement over the prototypical implementation presented in
\cite{Benkoe:03}. Besides the extention to the full chemical atom set GGL
implements an energy increment system as well as a rate calculation
approach for chemical reactions. Furthermore GGL provides an interface to the
OpenBabel library and therefore the whole functionality of this important open
source cheminformatics library can be harnessed from within the GGL.

\begin{figure}[tb]
  \begin{center}
    \includegraphics[width=\textwidth]{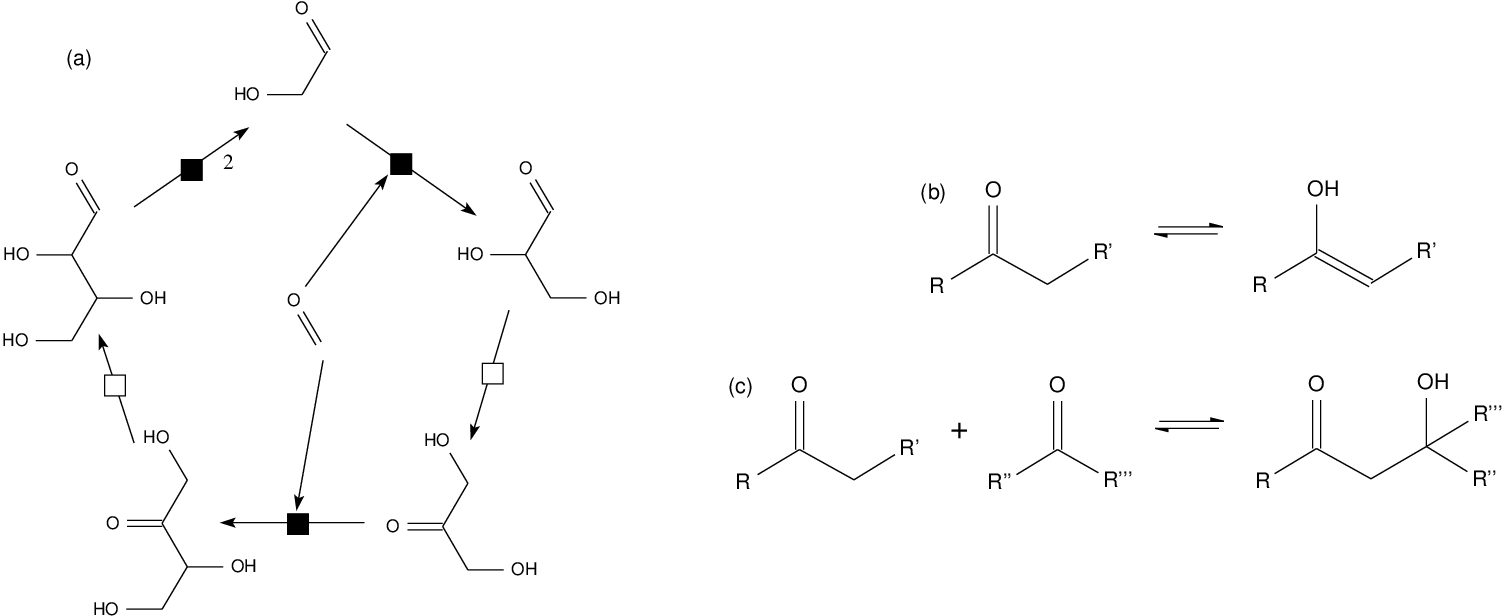}
  \end{center}
  \caption{Autocatalytic core network (a) of the formose process, keto-enol
    tautomerization of a carbonyl compounds (b), and aldol addition (c)
    between two carbonyl compounds. Depending on the reaction direction the
    aldol addition forms or breaks a \texttt{C}--\texttt{C} bond
    ($\blacksquare$). In the case of $\alpha$-hydroxy-carbonyl compounds
    (\texttt{HO}--\texttt{C}--\texttt{C}=\texttt{O}), the keto-enol
    tautomerization provides a mechanism for the carbonyl functionality
    (\texttt{C}=\texttt{O}) to shift along the backbone of the sugar
    ($\Box$).}
    \label{fig:formose-auto}
\end{figure}

\begin{figure}[tb]
  \begin{center}
     \includegraphics[width=\textwidth]{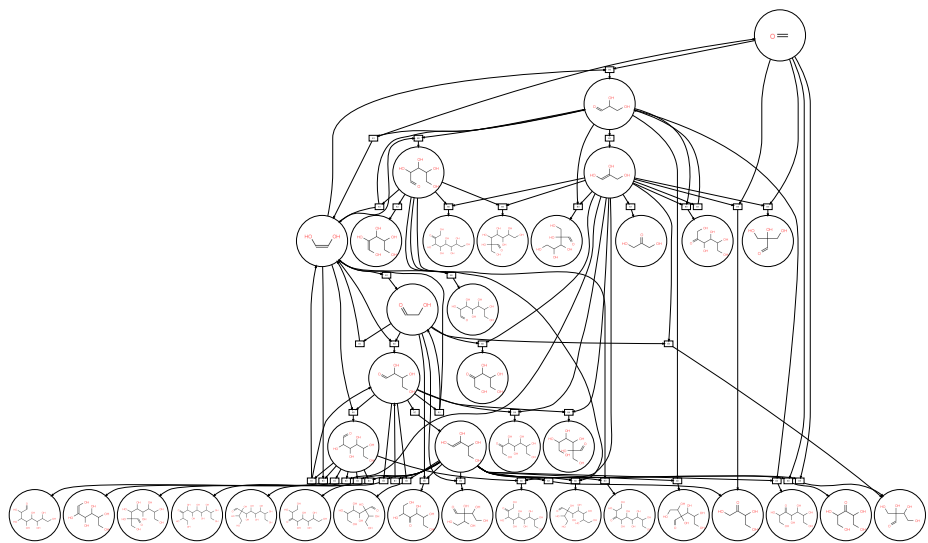}
    \caption{Hypergraph of the resulting reaction network after 4
      iterations of the formose process grammar.}
    \label{fig:formose-3iter}
  \end{center}
\end{figure}

On the basis of the formose process \cite{Butlerov:1861}, we illustrate,
that the graph grammar approach is indeed a sensible model of chemical
reaction networks and nicely captures the algebraic properties of chemistry
itself. The formose process condenses formaldehyde, the simplest possible
sugar, into a combinatorial complex mixture of higher sugars by repeatedly
involving only two reversible reactions, the aldol reaction and the
keto-enol tautomerization (see Fig.~\ref{fig:formose-auto}~b, c). The
corresponding graph grammar rule encodings are depicted in
Fig.~\ref{gml:keto-enol} and~\ref{gml:aldol} (suppl. material). An autocatalytic
loop, which produces glycolaldehyd and consumes formaldehyde, is located at the
core of the formose process (see Fig.~\ref{fig:formose-auto}~a).
Figure~\ref{fig:formose-3iter} depicts the growing reaction network of the
formose process.

The following table \ref{tab:formose-runtime} shows the exponential explosion of
the molecular space for the formose process and the according runtime of our
\toyChem{} tool. Note, the vast majority of the runtime is consumed by canonical
SMILES generation for the molecule graphs resulting from graph grammar rule
applications. Still this step is essential to distinguish new from already known
molecules and it is much faster than graph isomorphism based comparisons.

\begin{table}
\begin{center}
\begin{tabular}{r|>{\hfill}p{.1\textwidth}>{\hfill}p{.1\textwidth}>{\hfill}p{.1\textwidth}>{\hfill}p{.1\textwidth}>{\hfill}p{.1\textwidth}>{\hfill}p{.1\textwidth}p{1em}}
Iteration & 1 & 2 & 3 & 4 & 5 & 6 &\\
\hline
Molecules & 3 & 5 & 9 & 37 & 302 & 10,572 &\\
Time & 0 & 0 & 0 & 0 & 0.2$s$ & 10.5$s$ &
\end{tabular}
\end{center}
\caption{Exponential explosion of the molecular space for the formose process
starting from \texttt{OCC=O} and \texttt{C=O} along with the runtimes of
\toyChem{}.}
\label{tab:formose-runtime}
\end{table}

The second example shows the reaction network for the enzyme mechanism of
$\beta$-lactamase (see Fig.~\ref{fig:enzmechNet}) as found in the MACiE
database \cite{Holliday:2012} entry M0002. Specificity of the amino acide side
chains of Lys, Ser and Glu in the active site of the enzyme EC~3.5.2.6 was
achieved by marking the C$_\alpha$ atom labels of the aminoacides with a SMILES
class flag (see Fig.~\ref{gml:enzmechRule} suppl. material). In that way
reactions between amino acide side-chains are suppressed within the graph
grammar rule application. Figure~\ref{fig:enzmechMech} (suppl. material) depicts
the first step within the enzyme mechanism.

\section{Summary}

The Graph Grammar Library (GGL) is a powerful framework for cheminformatics
applications based on graph rewrite. It meets very well the mandatory
requirements for such studies as discussed by Yadav \etal{}~\cite{Yadav:04} and
comes with a powerful chemistry module providing essential algorithms. With
the advent of genome-scale metabolic networks, formalisms such as the GGL
to handle chemical transformation will become an important factor for the
analysis, interpretation, and manipulation of these networks.

%
%

\subsubsection*{Acknowledgements}
  
This work was supported in part by the Vienna Science and Technology Fund
(WWTF) proj. no. MA07-030, the Austrian GEN-AU project ``Bioinformatics
Integration Network III'', and the COST Action CM0703 ``Systems
Chemistry''. Furthermore, we thank the users of the GGL framework for their
ongoing contributions, bug reports and suggestions.
 
\bibliographystyle{splncs03}  
\bibliography{ggl-framework}      


\clearpage
\appendix
\begin{center}
\Large \bfseries \underline{Supplementary Material}
\end{center}

\renewcommand*\thefigure{S.\arabic{figure}}
\setcounter{figure}{0}

\section{Visualization}

\begin{figure}[h!] 
 
    \begin{center}
    	\includegraphics[width=0.6\textwidth]{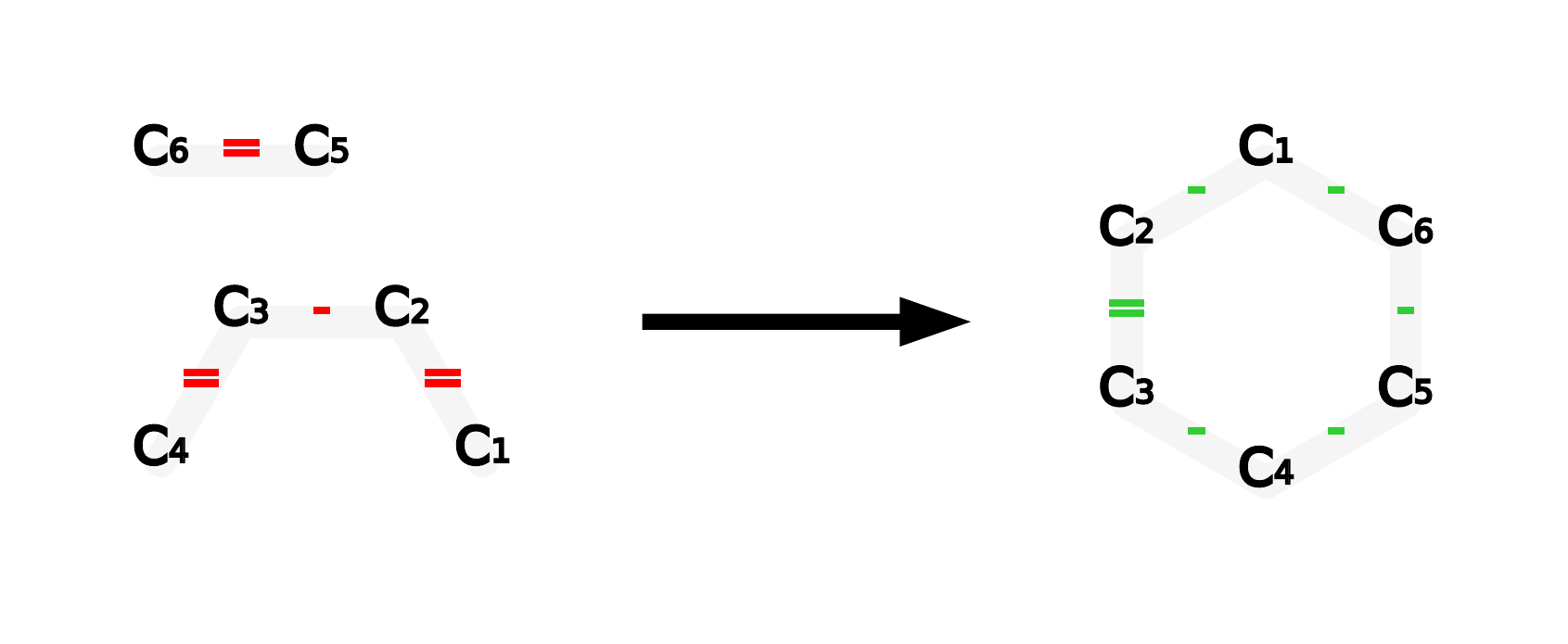}
    \end{center}
    \caption{The depiction of the Diels-Alder GML graph grammar rule encoding 
using the provided visualization script \chemruleSvg{}. The presentation highlights
the altered bonds within the left and right side graph of the reaction in red
and green, resp., to enable the identification of the underlying reaction
mechanism.}
	\label{fig:Diels-Alder}
\end{figure}

\FloatBarrier
\section{Molecular Groups}

\newcommand{\molcompsvg}{\texttt{molcomp2svg.pl}}

\def\NAD{NAD$^{+}$}
\def\NADH{NADH}

One field of application for the definition and use of molecular groups is the
specification of molecules that differ only in a few atoms or bonds. In such
cases, it can be convenient to specify only the dissimilar parts of the
molecules and to use group placeholders for the equal parts. That way, the
similarity becomes easy to see and the SMILES easier to read.

\begin{figure}[htb]
\begin{center}
  \includegraphics[width=0.45\textwidth]{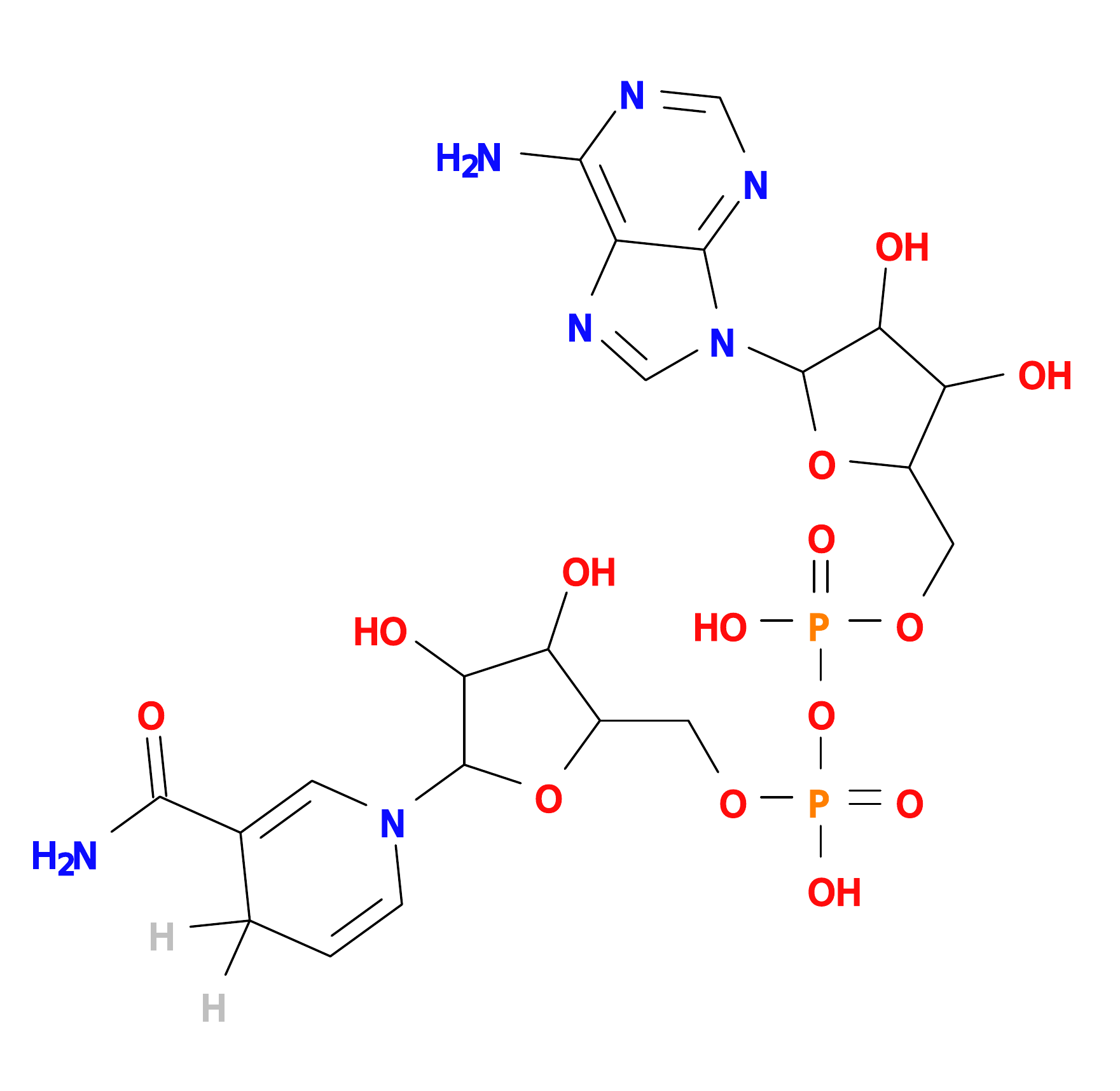}
  \includegraphics[width=0.45\textwidth]{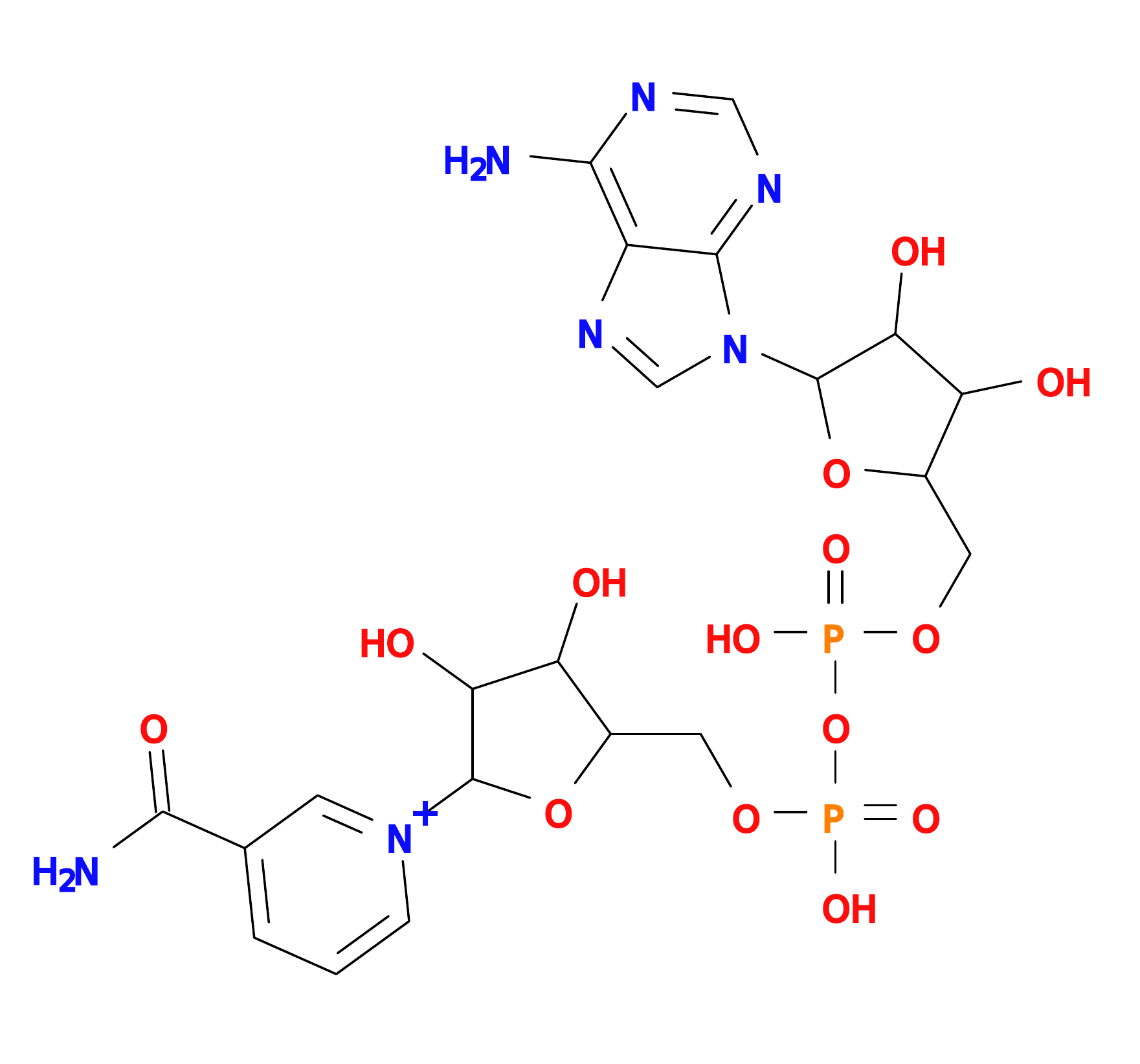}
  \caption{The molecules \NADH{} (left) and \NAD{} (right).}
  \label{fig:NADH}
\end{center}
\end{figure}

As an example, we use the molecules \NADH{} and \NAD{} depicted in
Fig.~\ref{fig:NADH} sporting 66 and 65 atoms, respectively. The difference
basically comprises only an additional proton within \NADH{} and a charge change
within the lower ring while the rest of the molecules are identically. Note, these two
changes alter the ring from non-aromatic (\NADH) to aromatic (\NAD).

Minimal SMILES encodings of the molecules (highlighting the differing ring in
red) are

\begin{center}
\begin{minipage}{0.9\textwidth}
\noindent\verb|NC(=O)|{\color{red}\verb|C1[CH2]C=CN(C=1|)}\verb|C2OC(COP(O)(=O)O..|\\
\noindent\verb|    ..P(O)(=O)OCC3OC(C(O)C3O)n4cnc5c(N)ncnc54)C(O)C2O|
\end{minipage}
\end{center}

\noindent for \NADH{} and 

\begin{center}
\begin{minipage}{0.9\textwidth}
\noindent\verb|NC(=O)|{\color{red}\verb|c1ccc[n+](c1)|}\verb|C2OC(COP(O)(=O)O..|\\
\noindent\verb|    ..P(O)(=O)OCC3OC(C(O)C3O)n4cnc5c(N)ncnc54)C(O)C2O|
\end{minipage}
\end{center}

\noindent for \NAD{}.

In contrast, when using group declarations for the identical parts, namely the
CONH2 group and the ribo-adenosine, the SMILES shrinks to 

\begin{center}
\begin{minipage}{0.9\textwidth}
\noindent\verb|[{CONH2}]|{\color{red}\verb|C1[CH2]C=CN(C=1)|}\verb|[{Ribo-ADP}]|
\end{minipage}
\end{center}

\noindent for \NADH{} and 

\begin{figure}[htb]
\begin{center}
  a)
  \includegraphics[height=0.25\textwidth]{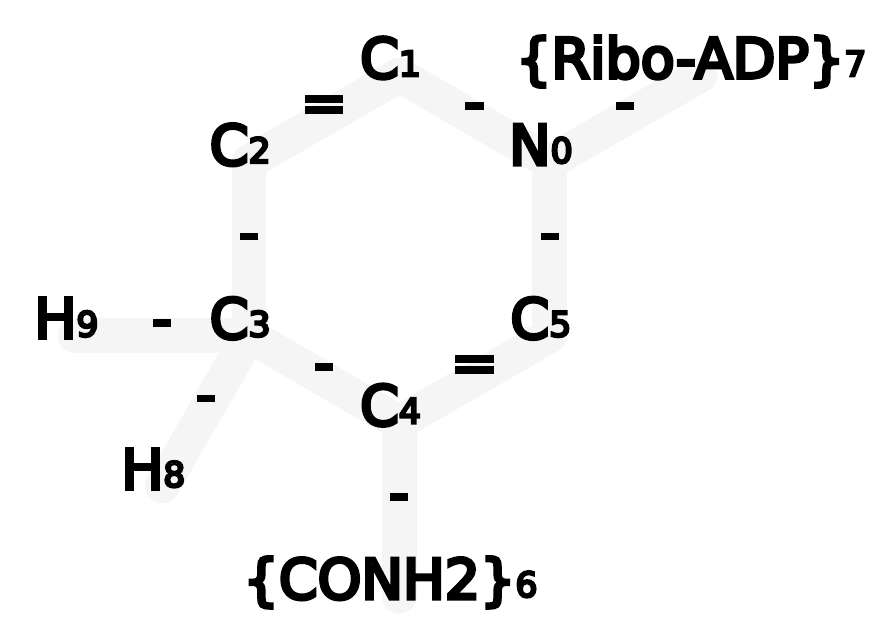}
  b)
  \includegraphics[height=0.25\textwidth]{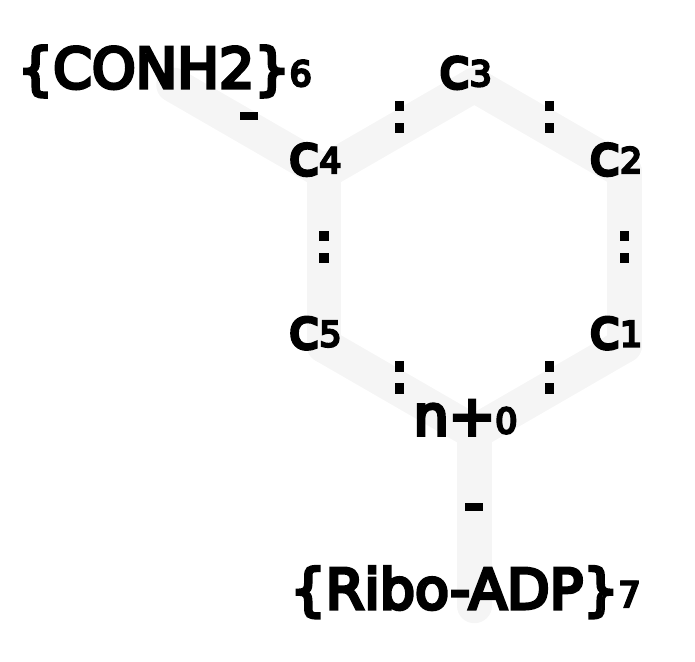}
  c)
  \includegraphics[height=0.2\textwidth]{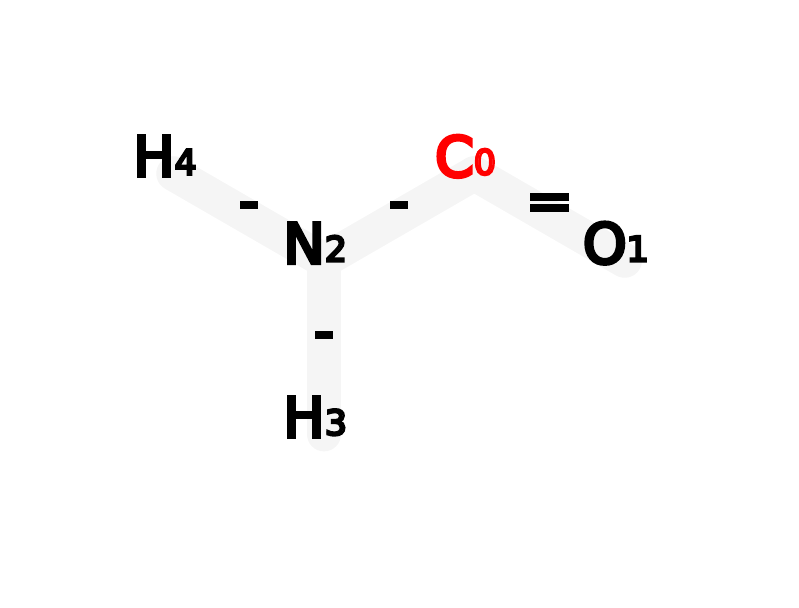}
  \caption{The molecules \NADH{} (a) and \NAD{} (b) described using
  group identifiers. Note, only for $C_3$ within \NADH{} explicit proton
  information is given; all other carbons have one not depicted adjacent
  proton. (c) Depiction of the \texttt{\{CONH2\}} group using the script
  \molcompsvg{}. Note, the proxy node is highlighted in red.}
  \label{fig:NADH-groups}
\end{center}
\end{figure}

\begin{center}
\begin{minipage}{0.9\textwidth}
\noindent\verb|[{CONH2}]|{\color{red}\verb|c1ccc[n+](c1)|}\verb|[{Ribo-ADP}]|
\end{minipage}
\end{center}

\noindent for \NAD{}, both depicted in Fig.~\ref{fig:NADH-groups}a) and~b).

\FloatBarrier

\section{General Graph Rewrite Examples}

\subsection{Game of Life}

In the 60th, John Conway created a cellular automaton named \emph{Game of Life}
\cite{Gardner:GameOfLife:70} that reflects the basic principles of birth, death,
and survival within populations. Each cell represents an individual that is
either alive or dead depending on the state of the neighbored individuals. 

The problem can be represented as a graph relabeling problem where each node
represents a cell and edges connect neighbored cells. The formulation of the
birth, death, and survival rules defined by Conway as graph grammar rules is
straight forwardly described by graph grammar rules. Each of the three rules
resembles a single node and according adjacency constraints for the matching and
the according recoloring of the node in match case. In combination with an
exhaustive application of the rules, this results in a complete Game of Life
solver based on a graph rewrite system distributed with the package.

\subsection{Sudoku}

Sudoku is a combinatorial problem where a given 9$\times$9 grid has to be
filled with numbers from~1 to~9, such that each number appears only once in each
row, column, and defined 3$\times$3 sub-grids. The task is to fill a given
partially filled grid such that all constraints are fullfilled
\cite{Simonis:05,Lynce:06}.

Similar to Game of Life, one can encode the Sudoku grid within a graph where
each cell is represented by a node. Each node is connected to all other cells
that have to have a different label. Thus, it resembles the dependency graph of
the problem. In order to find a solution, a Depths-First-Search (DFS) of the
exponential search space can be applied, where each search step is defined by
the valid application of a graph grammar rule. Each rule does the assignment of
a number to a non-assigned node while respecting all constraints.

The \GGLrelease{} package features a generic DFS implementation for such
purposes. Given a set of graph grammar rules and a start graph, a DFS
exploration of the search space is done and solutions are identified.

\subsection{Ring Perception}

The enumeration of all rings within a graph can be done using the algorithm
proposed by Hanser \etal{}~\cite{Hanser:96}. Such ring perceptions are important
e.g. in chemistry to do structure classifications or aromaticity identification
\cite{Roos-Kozel:81}. The algorithm by Hanser \etal{} creates an image of the
studied graph that represents the node adjacency within its edge labels, a so
called \emph{path graph}. This path graph is progressively collapsed in a way
that the final path graph contains only loops, each representing a ring
from the original graph.

We have implemented this general approach for ring perception based on a small
set of graph grammar rules and a simple rule application iteration. This shows
once more the expressivity of graph rewrite systems. A rule-based example as
well as an efficient native C++ implementation of the ring perception algorithm
is part of \GGLrelease{} package.

\subsection{\YD{}-equivalence problem}

The \YD{}-equivalence problem, also known by name wye-delta or delta-wye,
star-delta, star-mesh, or T-$\Pi$, is an important problem from graph theory
\cite{Archdeacon:98} with application in electrical resistor network
optimization \cite{Baran:97}. In short, two graphs are defined to be
\YD{}-equivalent if and only if they can be transformed into each other by
applying a series of \YD{}-transformations. These transformations are either to
convert a $\Delta$ triangle subgraph into a \texttt{Y}-like subgraph (by adding
a new central node in combination with the necessary rewiring) or the reverse
operation, i.e. transforming a \texttt{Y}-subgraph into a triangle (via center
deletion and rewiring). For instance, this property is fulfill by the Peterson
graph family forming a \YD{}-equivalence class \cite{Archdeacon:98}.

Using two simple graph grammar rules that encode the allowed
\YD{}-trans\-for\-ma\-tions, we can easily setup a search engine to check if two
graphs are \YD{}-equivalent or not. To this end, one starts two independent
Breadth-First-Searches (BFS) starting from the two graphs of interest. Within
each BFS all graphs are generated that can be obtained from the start graph by
applying \YD{}-transformations. The \YD{}-equivalence is proven as soon as the
two sets of graphs produced by the independent BFS intersect. The according
graph grammar rules in GML notation are given in Fig.~\ref{gml:wye-delta}.

\begin{figure}[htb]
\begin{center}
\begin{gml}
\begin{minipage}{0.7\textwidth}
\tiny
\makebox[\textwidth]{\hrulefill}
\begin{verbatim}
rule [ 
  ruleID "wye to delta" 
  wildcard "*" 
  context [ 
    node [ id 1 label "*" ] 
    node [ id 2 label "*" ] 
    node [ id 3 label "*" ] 
  ] 
  left [ 
    node [ id 4 label "*" ] 
    edge [ source 4 target 1 label "*" ] 
    edge [ source 4 target 2 label "*" ] 
    edge [ source 4 target 3 label "*" ]
    constrainAdj [ 
      id 4 op = count 3 
      edgeLabels [ label "*" ] 
      nodeLabels [ label "*" ] 
    ] 
    constrainNoEdge [ source 1 target 2 ] 
    constrainNoEdge [ source 1 target 3 ] 
    constrainNoEdge [ source 2 target 3 ]
  ]
  right [
    edge [ source 1 target 2 label "*" ] 
    edge [ source 1 target 3 label "*" ] 
    edge [ source 2 target 3 label "*" ]
  ] 
]
\end{verbatim}
\makebox[\textwidth]{\hrulefill}
\end{minipage}
\end{gml}
\end{center}
\caption{The GML rule encoding ``wye to delta'' graph grammar rule used for
  \YD{}~transformations. The reverse ``delta to wye'' rule can be easily
  obtained by exchanging the \texttt{left} and \texttt{right} content of
  the rule, while all constraints can be omitted.}
\label{gml:wye-delta}
\end{figure}

\FloatBarrier

\section{Chemical Reactions as Graph Grammars}

\subsection{Formose Process}

\begin{figure}[h!]
\begin{gml}
\begin{center}
\begin{minipage}{0.8\textwidth}
\tiny
\makebox[\textwidth]{\hrulefill}
\begin{verbatim}
rule [
  ruleID "Keto-Enol Isomerization"
  context [
    node [ id 1 label "C" ]
    node [ id 2 label "C" ]
    node [ id 3 label "O" ]
    node [ id 4 label "H" ]
  ]
  left [
    edge [ source 1 target 4 label "-" ]
    edge [ source 1 target 2 label "-" ]
    edge [ source 2 target 3 label "=" ]
    constrainAdj [ id 2 op = count 1 nodeLabels [ label "O" ] ]
  ]
  right [
    edge [ source 1 target 2 label "=" ]
    edge [ source 2 target 3 label "-" ]
    edge [ source 3 target 4 label "-" ]
  ]
]     
\end{verbatim}
\makebox[\textwidth]{\hrulefill}
\end{minipage}
\end{center}
\end{gml}
\caption{The GML rule encoding the ``keto-enol tautomerization'' graph
  grammar rule used. The reverse rule can be easily obtained by exchanging
  the \texttt{left} and \texttt{right} content of the rule, while all
  constraint can be omitted.}
\label{gml:keto-enol}
\end{figure}

\begin{figure}[h!]
\begin{gml}
\begin{center}
\begin{minipage}{0.8\textwidth} \tiny
\makebox[\textwidth]{\hrulefill}
\begin{verbatim}
rule [
  ruleID "Aldol Condensation"      
  context [
    node [ id 1 label "C" ]
    node [ id 2 label "C" ]
    node [ id 3 label "O" ]
    node [ id 4 label "H" ]
    node [ id 5 label "O" ]
    node [ id 6 label "C" ]
  ]
  left [
    edge [ source 1 target 2 label "=" ]
    edge [ source 2 target 3 label "-" ]
    edge [ source 3 target 4 label "-" ]
    edge [ source 5 target 6 label "=" ]
    constrainAdj [ id 2 op = count 1 nodeLabels [ label "O" ] ]
    constrainAdj [ id 6 op = count 1 nodeLabels [ label "O" ] ]
  ]
  right [
    edge [ source 1 target 2 label "-" ]
    edge [ source 2 target 3 label "=" ]
     edge [ source 5 target 6 label "-" ]
    edge [ source 4 target 5 label "-" ]
    edge [ source 6 target 1 label "-" ]
  ]       
]
\end{verbatim}
\makebox[\textwidth]{\hrulefill}
\end{minipage}
\end{center}
\end{gml}
\caption{The GML rule encoding the ``aldol condensation'' graph grammar
rule. The reverse rule for the retro-aldol can be easily obtained by
exchanging the \texttt{left} and \texttt{right} content of the rule,
while all constraint can be omitted.}
\label{gml:aldol}
\end{figure}

\clearpage
\subsection{$\beta$-Lactamase Enzyme Mechanism}

\begin{figure}[h!]
  \begin{center}
     \includegraphics[width=0.6\textwidth]{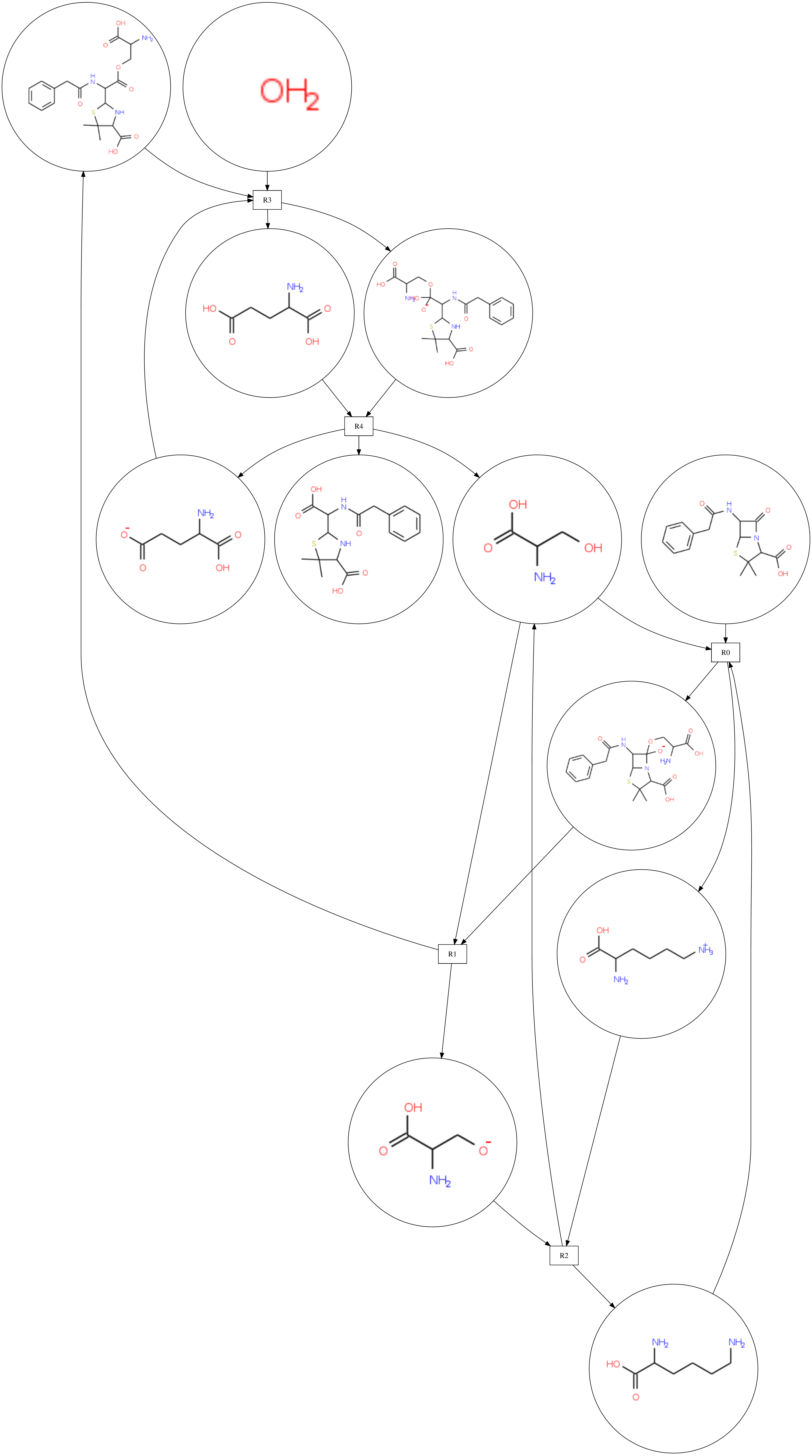}
    \caption{Hypergraph of the enzyme mechanism for $\beta$-lactamase (EC
      3.5.2.6).}
    \label{fig:enzmechNet}
  \end{center}
\end{figure}

\begin{figure}[h!]
    \begin{gml}
    \begin{center}
    \begin{minipage}{0.8\textwidth} \tiny
    \makebox[\textwidth]{\hrulefill}
\begin{verbatim}
# Beta-lactamase (class A)
# MACiE version: 3
# MACiE-entry: M0002, 3.5.2.6, Step 01
rule [
 ruleID "3.5.2.6-M0002-S01"
 left [
  node [ id 8 label "N" ]
  node [ id 5 label "O" ]
  edge [ source 1 target 5 label "=" ]
  edge [ source 6 target 7 label "-" ]
 ]
 context [
  node [ id 1 label "C" ]
  node [ id 2 label "C" ]
  node [ id 3 label "C" ]
  node [ id 4 label "N" ]
  node [ id 6 label "O" ]
  node [ id 7 label "H" ]
  node [ id 9 label "H" ]
  node [ id 10 label "H" ]
  node [ id 100 label "*"]
  node [ id 101 label "*"] 
  edge [ source 1 target 2 label "-" ]
  edge [ source 2 target 3 label "-" ]
  edge [ source 3 target 4 label "-" ]
  edge [ source 4 target 1 label "-" ]
  edge [ source 8 target 9 label "-" ]
  edge [ source 8 target 10 label "-" ]
  edge [ source 6 target 100 label "-" ]
  edge [ source 8 target 101 label "-" ]
  constrainNode [ id 100 op = nodeLabels [ label "C:1" ] ]
  constrainAdj [ id 100 op = count 1 nodeLabels [ label "O" ] ]
  constrainNode [ id 101 op = nodeLabels [ label "C:1" ] ] 
 ]
 right [
   node [ id 8 label "N+" ]
   node [ id 5 label "O-" ]
   edge [ source 1 target 5 label "-" ]
   edge [ source 1 target 6 label "-" ]
   edge [ source 8 target 7 label "-" ]
 ]
]
\end{verbatim}
\makebox[\textwidth]{\hrulefill}
    \end{minipage}
    \end{center}
    \end{gml}
    \caption{Rewrite Rule in the enzyme mechanism of $\beta$-lactamase (EC
      3.5.2.6). The distinction between enzyme and the substrate(s) is
      achieved by using \texttt{SMILES} class number for atom labels
      belonging to amino acid side chains of the enzyme.}
    \label{gml:enzmechRule}
\end{figure}

\begin{figure}[h!]
  \begin{center}
     \includegraphics[width=.9\textwidth]{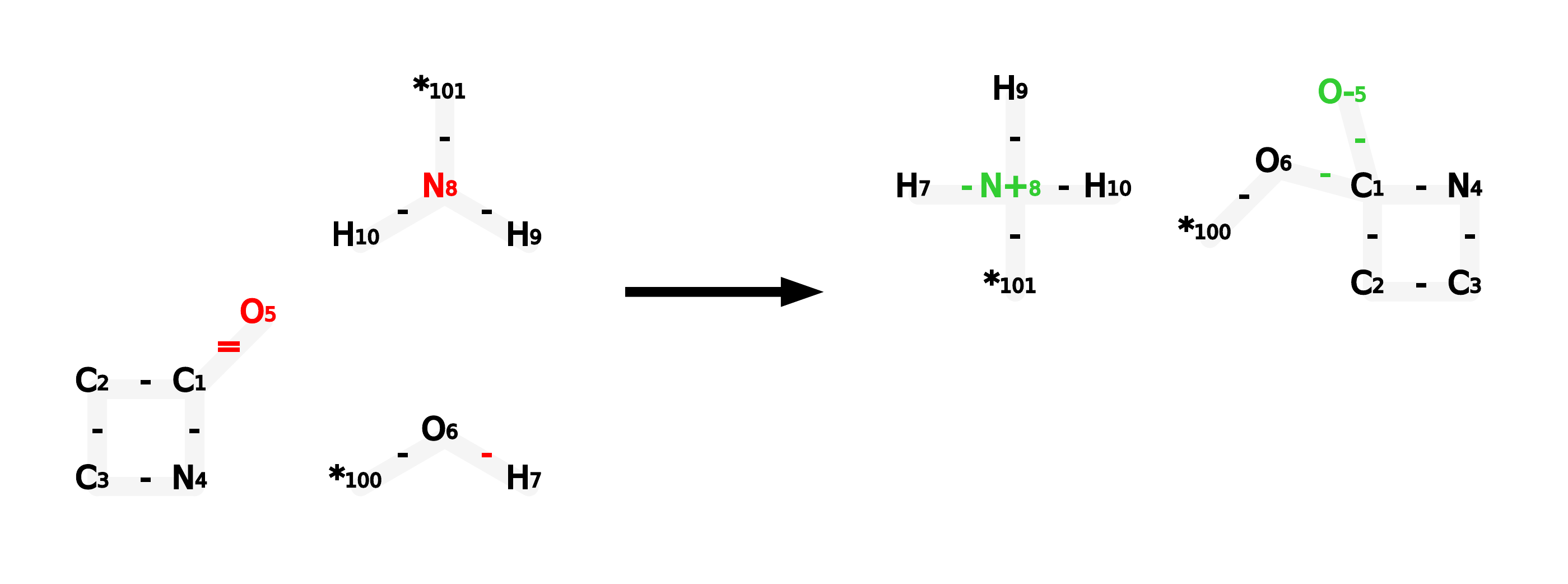}
  \end{center}
  \caption{First step in the enzyme mechanism of $\beta$-lactamase (EC 3.5.2.6)}
  \label{fig:enzmechMech}
\end{figure}

\end{document}